\documentclass[12pt,english,russian]{article}
\usepackage[T1]{fontenc}
\usepackage[koi8-r]{inputenc}

\makeatletter

\AtBeginDocument{\DeclareFontEncoding{T2A}{}{}}

\makeatletter

\AtBeginDocument{\DeclareFontEncoding{T2A}{}{}}

\makeatletter

\AtBeginDocument{\DeclareFontEncoding{T2A}{}{}}

\makeatletter

\AtBeginDocument{\DeclareFontEncoding{T2A}{}{}}

\makeatletter

\makeatletter

\makeatletter

\newtheorem{theorem}{Теорема}

\newtheorem{lemma}{Лемма}

\newtheorem{remark}{Remark}

\date{}

\usepackage{babel}

\begin{document}

\title{Critical states of strongly interacting many-particle systems on
a circle }

\author{V. A. Malyshev}

\maketitle

\section{Introduction}

In multicomponent systems with strong local interaction one can encounter
some phenomena absent in the standard systems of statistical physics
and other multicomponent systems. Namely, a system with $N$ components
in the bounded volume of order 1 (macroscale) has the natural microscale
of the order $\frac{1}{N}$. Applying the macroscopic force (of order
$1$) on the system, and thus on any of its components, one normally
gets changes on the macroscale itself and simultaneously small, of
the order $\frac{1}{N}$, changes of the microcomponents, see for
example \cite{Mal_Hooke}. In the systems, considered below, with
the strong Coulomb repulsion between the particles, however, one can
observe the influence of such force on the equilibrium state only
on a scale, much smaller that the standard microscale. Otherwise speaking,
the information about the macroforce is not available neither on the
macrocale nor on the standard microscale, but only on a finer scale.
If this phenomenon does not depend on the continuity properties of
the applied force, then the mere existence of the equilibrium depends
essentially on the continuity properties of the external force.

\paragraph{The model}

Consider the system \begin{equation}
0\leq x_{1}(t)<...<x_{N}(t)<L\label{X-periodic}\end{equation}
 of identical classical point particles on the interval $[0,L]$ with
periodic boundary conditions (that is on the circle $S$ of length
$L$). The dynamics of this system of points is defined by the system
of $N$ equations \begin{equation}
m\frac{d^{2}x_{i}}{dt^{2}}=-\frac{\partial U}{\partial x_{i}}+F(x_{i})-A\frac{dx_{i}}{dt}\label{main_eq}\end{equation}
 where $A\geq0$, $F$ is the external force, and the interaction
is given by

\[
U(x_{1},...,x_{N})=V(x_{2}-x_{1})+V(x_{3}-x_{2})+...+V(x_{1}-x_{N})\]
where $x_{1}-x_{N}$ should be understood as $x_{1}+(L-x_{N})$, otherwise
speaking, here and further on the differences are taken clockwise.
It is assumed that the potential $V$ is symmetric and repulsive,
and moreover\[
V(x)=V(-x)>0,V(r)=\alpha r^{-a+1}>0,r=|x|\]
 \begin{equation}
f(r)=-\frac{dV(r)}{dr}=\alpha(a-1)r^{-a}>0\label{powerLaw}\end{equation}
We assume further that $a>1$ and put $\alpha(a-1)=1$, that is $f(r)=r^{-a}$.
Fixed (critical) configurations $X=(x_{1},...,x_{N})$ of the system
are defined by the equations\begin{equation}
f(x_{k}-x_{k-1})+F(x_{k})-f(x_{k+1}-x_{k})=0,k=1,...,N\label{fixedEq}\end{equation}
assuming that positive forces are directed clockwise, and that $x_{0}=x_{N},x_{N+1}=x_{1}$.
Our goal will be to study these equations.

\paragraph{The results}

If $F\equiv0$, then it is evident that the fixed configuration is
unique up to translation, and for any $i=1,...,N$ \[
|x_{i+1}-x_{i}|=\frac{L}{N}\]
If $F$ is not identically zero, the situation is essentially more
complicated. However, we have the following general result.

\begin{theorem}

Let the external force $F(x)$ be a bounded function. Assume that
there exists a sequence $(x_{1}^{(p)},...,x_{N_{p}}^{(p)}),p=1,2,...,$
of fixed configurations with $N_{p}\to\infty$. Then for $p\to\infty$
uniformly in $i=1,...,N_{p}$ \[
|x_{i+1}^{(p)}-x_{i}^{(p)}|\sim\frac{L}{N_{p}}\]

\end{theorem}

Existence and uniqueness of the fixed configuration do not have such
general general results, but have interesting effects, related to
$N\to\infty$. Let the force be potential on the circle, that is\[
\int_{S}F(x)dx=0\]
then the potential can be defined as \[
W(x)=\int_{0}^{x}F(x)dx\]
If $W(x)$ is smooth, then the minimum of the potential\[
U(x_{1},...,x_{N})+\sum_{i=1}^{N}W(x_{i})\]
satisfies the equation (\ref{fixedEq}), this minimum thus being the
solution of these equations. At the same time we have

\begin{lemma}

Let $F(x)$ be left-continuous function with finite number of gaps.
If there exists a sequence $N_{p}\to\infty$ such that for any $p$
there exists at least one fixed configuration $(x_{1}^{(p)},...,x_{N_{p}}^{(p)})$,
then\begin{equation}
\int_{S}Fdx=0\label{cycle}\end{equation}
 that is the force $F(x)$ is potential on the circle.

\end{lemma}

The proof of this lemma will be obtained during the proof of lemma
2.

If the potential $W(x)$ is not smooth, even if the force is potential,
the equations (\ref{fixedEq}) are far from being always solvable,
that is critical points cannot exist at all. This is not surprising,
similar phenomenon exists for example in the following simplest model
with one particle, where there is no critical point on the interval
$[0,L]$: the particle moves in the field of the external force $F(x)=F_{1}>0,x\in[0,M],0<M<L,$
and $F(x)=F_{2}<0,x\in(M,L]$. However we concentrate only on phenomena
related to the number of particles $N\to\infty$

Even the simplest case of the piecewise constant function shows that
the existence of the equilibrium is not generic. Let $F(x)=F_{1}$
on the interval $(0,M]$ of the circle $S$, $F(x)=F_{2}$ on its
complement $(M,L]$. We shall often denote $M_{1}=M,M_{2}=L-M$, $N_{1},N_{2}$
- the number of particles on the intervals $(0,M]$ and $(M,L]$ correspondingly,
$N=N_{1}+N_{2}$.

Our goal is to prove the following results.

\begin{theorem}

Fix $L,M=\frac{L}{2}$ and $F(x)=F>0,x\in(0,M],F(x)=-F,x\in(M,L]$.
Then for sufficiently large even $N$ there is a continuum of fixed
configurations. More exactly, for any point $x\in S$ there is a fixed
configuration\[
0<x_{1}<...<x_{\frac{N}{2}}\leq\frac{M}{2}<x_{\frac{N}{2}+1},...,x_{N}\leq L,\]
containing $x$ (that is $x$ coinsides with some $x_{k}$). Moreover,
the number of points on the intervals $(0,M]$ and $(M,L]$ is the
same, and $\Delta_{k}=x_{k+1}-x_{k}=\Delta_{N-k}$ for all $k=1,...,\frac{N}{2}$.

If $N$ is odd, there is no fixed configurations.

\end{theorem}

During the proof we shall explicitely construct the existing fixed
configurations.

\begin{theorem}

For any $0<C_{1}<C_{2}<\infty$ and any sequence $N_{1}^{(p)},N_{2}^{(p)}\to\infty$
so that $\frac{N_{1}^{(p)}}{N_{2}^{(p)}}\to\gamma\neq1$, starting
from some $p$ there is no fixed configurations for any $L,M,F_{i}$
such that\[
C_{1}<L,M,F_{i}<C_{2}\]

At the same time one can always change the value of the  function
$F$ at one or two gap points (that is at the points $M$ or $L$)
so, that there exists a fixed configuration with one or two particles
at the gap points.

\end{theorem}

\section{Proofs}

\subsection{Uniform asymptotics}

Let us prove theorem 1. For any $p$ there exists at least one $1\leq k(p)\leq N$
with\[
\Delta_{k(p)}^{(p)}=x_{k(p)+1}^{(p)}-x_{k(p)}^{(p)}\leq\frac{L}{N_{p}}\]
Thus there can be two cases. Either for $p\to\infty$\[
\Delta_{k(p)}^{(p)}\sim\frac{L}{N_{p}}\]
either there is a subsequence $p_{n}$ such that for any $\epsilon>0$
and all $p_{n}$\[
\Delta_{k(p_{n})}^{(p_{n})}\leq\frac{L}{N_{p_{n}}}(1-\epsilon)\]
Let us prove that in the first case the theorem holds. In fact, let
us sum up the equations (\ref{fixedEq}) with $k=k(p)+1,...,m$. Here
$m$ can be any number of\[
m=k(p)+1,...,N,1,...,k(p)-1\]
 (we consider the indices modulo $N$, that is we identify $x_{k}$
and $x_{k+N}$ for any integer $k$). Then\[
f(\Delta_{m}^{(p)})=f(\Delta_{k(p)}^{(p)})+F(x_{k(p)+1}^{(p)})+...+F(x_{m}^{(p)})\]
 and thus for some $C=\sup|F(x)|>0$\[
f(\Delta_{m}^{(p)})=f(\Delta_{k(p)}^{(p)})+r_{m}^{(p)},|r_{m}^{(p)}|\leq CN_{p}\]
 or\[
\Delta_{m}^{(p)}=((\Delta_{k(p)}^{(p)})^{-a}+r_{m}^{(p)})^{-\frac{1}{a}}\sim\frac{L}{N_{p}}(1+r_{m}^{(p)}\frac{L^{a}}{N_{p}^{a}})^{-\frac{1}{a}}\]
The result follows from this. Let us prove now that the second case
is impossible. Quite similarly, for any sufficiently large $p_{n}$
and all $m$ \[
\Delta_{m}^{(p_{n})}\leq(1-\epsilon)\frac{L}{N_{p_{n}}}(1+r_{m}^{(p_{n})}\frac{L^{a}(1-\epsilon)}{N_{p_{n}}^{a}})^{-\frac{1}{a}}\leq(1-\frac{\epsilon}{2})\frac{L}{N_{p_{n}}}\]
From this, summing over по $m=1,...,N$, we get $L\leq(1-\frac{\epsilon}{2})L$,
which is impossible.

\subsection{Existence conditions}

The following lemma (together with lemma 1) gives a list of obstructions
for the existence of the fixed points.

\begin{lemma} 
\begin{itemize}
\item If the quotient $\frac{F_{2}}{F_{1}}$ is irrational, then there are
no fixed points. 
\item For fixed $M_{i},F_{i},N$ the fixed configuration can exist not more
than for one partition of the number $N=N_{1}+N_{2}$, where $N_{i}$
is the number of particles on the interval of length $M_{i}$ correspondingly. 
\item Let $M_{i},F_{i}$ be fixed. Then a necessary condition of existence
of at least one fixed configuration for any $p$ in a sequence $N^{(p)},p=1,2...,$
is the existence of the partition $N_{1}^{(p)}+N_{2}^{(p)}=N$ such
that\begin{equation}
\frac{M_{1}}{M_{2}}=-\frac{F_{2}}{F_{1}}=\frac{N_{1}^{(p)}}{N_{2}^{(p)}}\label{cond_3}\end{equation}

\end{itemize}
\end{lemma}

Proof. Note that for fixed $N$ a necessary condition for the configuration
to be fixed is the condition\begin{equation}
\sum F(x_{i})=0\label{cycle_N}\end{equation}
which is obtained by summing the equations (\ref{fixedEq}). In our
case (\ref{cycle_N}) becomes \begin{equation}
F_{1}N_{1}+F_{2}N_{2}=0\label{sily}\end{equation}
From this the two first assertions of the lemma follow. Let us note
that there appears the necessary condition of the arithmetic character:
$F_{1}$ and $F_{2}$ should be commensurable.

Note now that by theorem 1 for $p\to\infty$\[
|\frac{L}{N_{p}}\sum_{i=1}^{N_{p}}F(x_{i}^{(p)})-\sum_{i=1}^{N_{p}}F(x_{i}^{(p)})(x_{i+1}^{(p)}-x_{i}^{(p)})|\to0\]
 and thus \[
\frac{L}{N_{p}}\sum_{i=1}^{N_{p}}F(x_{i}^{(p)})\to_{p\to\infty}\int_{S}Fdx=F_{1}M_{1}+F_{2}M_{2}\]
 which gives the proof of lemma 1, as by (\ref{cycle_N}) the left
part is identically zero. Besides this, we get the first equality
in (\ref{cond_3}), and the second follows from (\ref{sily}).

\subsection{Auxiliary problem on the segment}

Consider the system of identical classical point particles on the
interval $[0,L]\in R$ \[
0\leq x_{1}<...<x_{N}\leq L\]
with the same interaction (\ref{powerLaw}) but with the completely
inelastic boundary conditions. This means that if one of the extreme
particles reaches one of the end points of the interval, it stops
and can leave this point only if the resulting force becomes directed
to inside the interval. Then the equilibrium condition is the following
system\[
F(0)-f(x_{2})\leq0,F(L)+f(L-x_{N-1})\geq0\]
 \begin{equation}
f(x_{k}-x_{k-1})+F(x_{k})-f(x_{k+1}-x_{k})=0,k=2,...,N-1\label{intervalEqu}\end{equation}
We will consider the fixed configurations such that $x_{1}=0,x_{N}=L$.
One can show that there are no others but we will not need this.

\begin{lemma}Assume that the external force $F>0$ is constant. Then
for sufficiently large $N$ the fixed point $(x_{1},x_{2},...,x_{N})$
such that $x_{1}=0,x_{N}=L$, exists and is unique, moreover $\Delta_{k}=x_{k+1}-x_{k},k=2,...,N-1$,
analytically depends on $L$, $F$ and $\Delta_{1}$.

\end{lemma}

Proof. It is convenient to introduce $\delta_{k}$ by \[
\Delta_{k}=x_{k+1}-x_{k}=\frac{L}{N-1}(1+\delta_{k})\]
 We have from (\ref{intervalEqu}) \begin{equation}
f(x_{k+1}-x_{k})=f(x_{2})+(k-1)F\label{intervalEqu_2}\end{equation}
 or\[
f(\frac{L}{N-1}(1+\delta_{k}))-f(\frac{L}{N-1}(1+\delta_{1}))=(k-1)F\]
 Rewrite\begin{equation}
(1+\delta_{k})^{-a}-(1+\delta_{1})^{-a}=Q_{k}=(k-1)q,q=(\frac{L}{N-1})^{a}F\label{delta_equations}\end{equation}
 or\begin{equation}
1+\delta_{k}=[(1+\delta_{1})^{-a}+Q_{k}]^{-\frac{1}{a}}\label{delta_k}\end{equation}
that defines $\delta_{k}$ as the real analytic function of $L,F,\delta_{1}$
for $L>0,F>0,|\delta_{1}|<1$. One can write

\[
\delta_{k}=[1-a\delta_{1}+\frac{a(a+1)}{2}\delta_{1}^{2}+O(\delta_{1}^{3})+Q_{k}]^{-\frac{1}{a}}-1=\]
 \begin{equation}
=\delta_{1}-a^{-1}Q_{k}+\frac{a^{-1}(a^{-1}+1)}{2}Q_{k}^{2}-(a^{-1}+1)\delta_{1}Q_{k}+g_{3}(\delta_{1},N,k)\label{delta}\end{equation}
 where \[
g_{3}(\delta_{1},N,k)=O((|\delta_{1}|+Q_{k})^{3})\]
 is the real analytic function of $L,F,\delta_{1}$ for $L>0,F>0,|\delta_{1}|<1$.

Summing over $k$ and using the condition\[
\sum_{k=1}^{N-1}\delta_{k}=0\]
 we get\[
\sum_{k=1}^{N-1}\delta_{k}=0=(N-1)\delta_{1}-(a^{-1}+(a^{-1}+1)\delta_{1})\sum_{k=2}^{N-1}Q_{k}+\frac{a^{-1}(a^{-1}+1)}{2}\sum_{k=2}^{N-1}Q_{k}^{2}\]
 \[
+\sum_{k=2}^{N-1}g_{3}(\delta_{1},N,k)\]
 It follows\[
\delta_{1}=(N-1)^{-1}((a^{-1}+(a^{-1}+1)\delta_{1})\sum_{k=2}^{N-1}Q_{k}-\frac{a^{-1}(a^{-1}+1)}{2}\sum_{k=2}^{N-1}Q_{k}^{2})+G(\delta_{1},N)=\]
 \[
=(a^{-1}+(a^{-1}+1)\delta_{1})q(\frac{N}{2}-1)-\frac{a^{-1}(a^{-1}+1)}{2}q^{2}\frac{(N-2)(2N-3)}{6}+G(\delta_{1},N)\]
 \[
G(\delta_{1},N)=(N-1)^{-1}\sum_{k=2}^{N-1}g_{3}(\delta_{1},N,k)=O((|\delta_{1}|+Q_{N-1})^{3})=O((|\delta_{1}|+N^{-a+1})^{3})\]
As the latter equation can be written as $\delta_{1}=\sum_{n=0}^{\infty}c_{n}\delta_{1}^{n}$,
where in the righthand part there is an analytic function with small
coefficients $c_{n}$, then by subsequent iterations we get the unique
solution for $\delta_{1}$. Moreover\begin{equation}
\delta_{1}=a^{-1}q(\frac{N}{2}-1)+\frac{a^{-1}(a^{-1}+1)}{12}q^{2}(N-2)(N-3)+J_{1}(N),J_{1}(N)=o(N^{-2a+2})\label{Exp_delta1}\end{equation}
 \[
\delta_{N-1}=\delta_{1}-a^{-1}Q_{N-1}+\frac{a^{-1}(a^{-1}+1)}{2}Q_{N-1}^{2}-(a^{-1}+1)\delta_{1}Q_{N-1}+g_{3}(\delta_{1},N,N-1)=\]
 \begin{equation}
=-a^{-1}q(\frac{N}{2}-1)+\frac{a^{-1}(a^{-1}+1)}{12}q^{2}(N-2)(N-3)+J_{N-1}(N),J_{N-1}(N)=o(N^{-2a+2})\label{Exp_delta_N-1}\end{equation}

\subsection{Construction of fixed configurations}

Note that if $X=(x_{1},...,x_{N})$ is a fixed configuration on the
circle with given $F(x)$, then any its subsequence (without gaps)
$X_{kl}=(x_{k},...,x_{l})$ is a fixed configuration on the segment
$[x_{k},x_{l}]$ (with the same force $F(x)$) in the sense of the
section 2.3. In fact, restricting on the segment we neglect the part
of the force at the end points, directed to inside the interval. It
means that at the end points the force becomes directed to outside
the segment, which gives first two inequalities in (\ref{intervalEqu}).
Vice-versa, if for a given configuration $X=(x_{1},...,x_{N})$ $X_{kl}=(x_{k},...,x_{l})$
will be the fixed configuration (with the same force $F(x)$) on the
segment of the circle in-between the points $x_{k}$ and $x_{l}$
(in the clockwise order), and the configuration $X_{kl}=(x_{l-1},...,x_{k+1})$
will be the fixed configuration on the segment in-between the points
$x_{l-1}$ and $x_{k+1}$, then $X$ is the fixed configuration on
the circle. Such situation is called glueing. In the symmetric case,
that is under the conditions of theorem 2, in glueing we use the mirror
symmetry.

\paragraph{Symmetric case}

Let us prove theorem 2. Let $N$ be even and $x\in S$. We will construct
the fixed configuration, containing $x$, for the case of equal number
$\frac{N}{2}$ of points on the intervals $(0,M]$ and $(M,L]$.

Let us consider the fixed configuration with $\frac{N}{2}+2$ points\[
0=y_{1}<...<y_{\frac{N}{2}+2}=M+m\]
on the interval $[0,M+m]$ ($m>0$ being a small real number) with
constant force $F>0$, as in the section 2.3. By lemma 3 it is unique
and has the differences \[
\Delta_{k}=y_{k+1}-y_{k}=\frac{M+m}{\frac{N}{2}+1}(1+\delta_{k}),k=1,...,\frac{N}{2}+1\]
 \[
\sum_{k=1}^{\frac{N}{2}+1}\Delta_{k}=M+m\]
which were calculated in the section 2.3. With these differences we
will construct a fixed configuration $X=(x_{1},...,x_{N})$ on the
circle. Define by clockwise induction, for some $b>0$,

\[
x_{N}=L-b,x_{1}=-b+\Delta_{1},x_{2}=x_{1}+\Delta_{2},...,x_{\frac{N}{2}+1}=x_{\frac{N}{2}}+\Delta_{\frac{N}{2}+1}=-b+\sum_{k=1}^{\frac{N}{2}+1}\Delta_{k}=-b+M+m\]
 and similarly by counter-clockwise induction \[
x_{N}=L+x_{1}-\Delta_{1},x_{N-1}=x_{N}-\Delta_{2},...,x_{N-k}=x_{N-k+1}-\Delta_{k+1},...,x_{\frac{N}{2}}=x_{\frac{N}{2}+1}-\Delta_{\frac{N}{2}+1}\]
 We see that these definitions are compatible and it follows from
them\[
x_{\frac{N}{2}+1}-x_{1}=x_{N}-x_{\frac{N}{2}}=M+m-\Delta_{1}\]
For the constructed configuration to be a configuration on the circle
of length $L$, it is necessary the additional condition $2(M+m)-\Delta_{1}-\Delta_{\frac{N}{2}+1}=L$
(as two intervals of length $M+m$ cover the circle, but the intervals
$\Delta_{1},\Delta_{\frac{N}{2}+1}$ are taken into account twice)
or \begin{equation}
2m=\Delta_{1}+\Delta_{\frac{N}{2}+1}\label{length}\end{equation}
 This gives the equation for $m$\[
2m=\frac{M+m}{\frac{N}{2}+1}(1+\delta_{1})+\frac{M+m}{\frac{N}{2}+1}(1+\delta_{\frac{N}{2}+1})=\frac{M}{N}+h(m,N)\]
where the function $h$ is analytic in $m$ and by following the expansions
(\ref{Exp_delta1},\ref{Exp_delta_N-1}) \[
h(m,N)=o(\frac{|m|}{N}+\frac{1}{N^{2}})\]
That is why this equation has a unique solution $m=O(\frac{1}{N})$.

From the definitions above it follows that $\Delta_{k}=\Delta_{k}(m)$
depends on $m$ and $x_{k}=x_{k}(m,b)$ depends on $m$ and $b$,
and moreover, as we know fron section 2.3, for all $k$ $\Delta_{k}(m)>\Delta_{k+1}(m)$.
In particular, for given $m$, $\Delta_{1}(m)$ is the maximal of
the intervals $\Delta_{k}(m)$. Besides that, \[
\Delta_{k}(m')>\Delta_{k}(m),m'>m\]
 \[
x_{k}(m,b+c)=x_{k}(m,b)+c\]

To get a fixed point on the circle from this glueing (that is to satisfy
equilibrium conditions), one should demand that the points $x_{1},...,x_{\frac{N}{2}}$
belonged to the interval $(0,M]$, and the rest belonged to the interval
$(M,L]$. Necessary and sufficient conditions for this will be the
inequalities\begin{equation}
x_{1}>0\label{mirror_1}\end{equation}
\begin{equation}
x_{\frac{N}{2}}=-b+M+m-\Delta_{\frac{N}{2}+1}<M\label{mirror_2}\end{equation}
\begin{equation}
x_{\frac{N}{2}+1}=-b+M+m>M\label{mirror_3}\end{equation}
 The first one can be reduced to\begin{equation}
0<b<\Delta_{1}(m)\label{mirror_4}\end{equation}
 The second and the third ones can be reduced to\begin{equation}
b<m<\Delta_{\frac{N}{2}+1}(m)+b\label{mirror_5}\end{equation}
 Put\[
b(m)=\frac{\Delta_{1}(m)}{2}\]
It is easy to see that all inequalities (\ref{mirror_4},\ref{mirror_5})
are fullfilled.

Assume that $x_{k}(m,b(m))$ is the point of the fixed configuration,
with the parameters $m,b(m)$, which is the nearest to $x$ . Let
for example $x_{k}<x$. Then\[
x-x_{k}\leq\frac{\Delta_{k-1}(m)}{2}\]
Choosing now $b=b(m)+x-x_{k}$, we get the point $x$ as the $k$-th
point of the new configuration\[
x=x_{k}(m,b)\]
Nonexistence for odd $N$ follows from the third assertion of lemma
2.

\begin{remark}

For given $x$ and $k$, the fixed configuration with $x=x_{k}$ is
unique, which follows from the monotonicity of the function $x_{k}(b)$
in $b$. The question whether there can be, for given $x$, two fixed
configurations such that for one of them $x=x_{k}$, and for the other
one $x=x_{k+1}$, acquires more exact calculations and is not considered
here.

\end{remark}

\paragraph{Asymmetric case}

Let us prove theorem 3. Let for some $p$ there exist a fixed configuration,
further on we omit the index $p$. Then it looks like\[
0<x_{1}<x_{2}<...<x_{N_{1}}\leq M<x_{N_{1}+1}<...<x_{N}\leq L\]
that is the points $x_{1},...,x_{N_{1}+1}$ belong to the interval
where the force $F_{1}>0$ is applied, and the points $x_{N_{1}+1},...,x_{N}$
belong to the interval where the force $F_{2}<0$ is applied. This
configuration defines two auxiliary fixed configurations on the intervals
$[0,M_{i}+m_{i}],i=1,2,$ with $N_{i}+2$ points\[
0=y_{1}^{i}<...<y_{N_{1}+2}^{i}=M_{i}+m_{i}\]
 and the forces $F_{i}$ correspondingly, which are defined by their
differences \[
u_{i,k}=y_{k+1}^{i}-y_{k}^{i}=\frac{M_{i}+m_{i}}{N_{i}+1}(1+\delta_{i,k})\]
Moreover, $u_{1,k}$ are defined by the coordinates $x_{N},x_{1},...,x_{N_{1}+1}$

\[
x_{1}=L-x_{N}+u_{1,1},x_{2}=x_{1}+u_{1,2},...,x_{N_{1}+1}=x_{N_{1}}+u_{1,N_{1}+1}\]
 and $u_{2,k}$ is defined by the coordinates $x_{1},x_{N},x_{N-1},...,x_{N_{1}}$
(in reverse order)\[
x_{N}=x_{1}+L-u_{2,1},x_{N-1}=x_{N}-u_{2,2},...,x_{N_{1}}=x_{N_{1}+1}-u_{2,N_{2}+1}\]
 For compatibility the following two conditions should be fullfilled
\begin{equation}
u_{1,1}=u_{2,1},u_{1,N_{1}+1}=u_{2,N_{2}+1}\label{intervaly}\end{equation}
and the length $L$ will be defined by\begin{equation}
L=M_{1}+m_{1}+M_{2}+m_{2}-u_{1,1}-u_{1,N_{1}+1}\label{summa}\end{equation}
that is simalarly to (\ref{length}) as the circle of length $L$
is covered by two segments, where $u_{1,1}$ and $u_{N_{1}+1}$ in
the union of two segments are counted twice. Also it should be $F_{2}N_{2}+F_{1}N_{1}=0$.
It is convenient to denote $\hat{M}_{i}=M_{i}+m_{i}$.

As (by lemma 2 or by theorem 1) \begin{equation}
\frac{\hat{M}_{1}}{N_{1}+1}\sim\frac{\hat{M}_{2}}{N_{2}+1}\label{M1_M_2}\end{equation}
 then one can write\begin{equation}
\frac{\hat{M}_{1}+m}{N_{1}+1}=\frac{\hat{M}_{2}}{N_{2}+1},m=o(1)\label{M1_M2_m}\end{equation}
and find $m$ from equation (\ref{intervaly}), that will give two
equations for $m$ \[
\frac{\hat{M}_{1}}{N_{1}+1}(1+\delta_{1,1})=\frac{\hat{M}_{1}+m}{N_{1}+1}(1+\delta_{2,1})\]
\[
\frac{\hat{M}_{1}}{N_{1}+1}(1+\delta_{1,N_{1}+1})=\frac{\hat{M}_{1}+m}{N_{1}+1}(1+\delta_{2,N_{2}+1})\]
 Rewrite the latter equations as\begin{equation}
m=\hat{M}_{1}(\delta_{1,1}-\delta_{2,1})(1+\delta_{2,1})^{-1}\label{m_eq_1}\end{equation}
 \begin{equation}
m=\hat{M}_{1}(\delta_{1,N_{1}+1}-\delta_{2,N_{2}+1})(1+\delta_{2,N_{2}+1})^{-1}\label{m_eq_N-1}\end{equation}
We obtained two equations with one unknown $m$. Let us show that
they are incompatible for small $m$. To do this let us first compare
the main terms of the two series: for $\delta_{1,1}-\delta_{2,1}$
and $\delta_{1,N_{1}+1}-\delta_{2,N_{2}+1}$. More exactly, using
the expansion (\ref{Exp_delta1}), calculate the difference of the
two first terms in the series for $\delta_{1,1}$ and $\delta_{2,1}$.
Note that, doing this, we will use $|F_{2}|N_{2}=F_{1}N_{1}$, and
in the formulae for $\delta_{2,1}$ we should take $|F_{2}|$, as
$y_{k}^{2}$ corresponds to the inverse order of the coordinates $x_{i}$.
Direct calculation gives\[
\delta_{1,1}-\delta_{2,1}=R_{1}+R_{2}\]
 where\[
R_{1}=\frac{1}{2}a^{-1}F_{1}N_{1}(\frac{\hat{M}_{1}}{N_{1}+1})^{a}[1-(1+\frac{m}{\hat{M}_{1}})^{a}]=\frac{1}{2}a^{-1}F_{1}N_{1}(\frac{\hat{M}_{1}}{N_{1}+1})^{a}[-a\frac{m}{\hat{M}_{1}}+O(m^{2})]\]
 \[
R_{2}=\frac{a^{-1}(a^{-1}+1)}{12}F_{1}N_{1}(\frac{\hat{M}_{1}}{N_{1}+1})^{2a}[N_{1}-(1+\frac{m}{\hat{M}_{1}})^{2a}(N_{2}-1)]=\]
 \[
=\frac{a^{-1}(a^{-1}+1)}{12}F_{1}N_{1}(\frac{\hat{M}_{1}}{N_{1}+1})^{2a}[(N_{1}-N_{2})-(2a\frac{m}{\hat{M}_{1}}+O(m^{2}))(N_{2}-1)]\]
Thus the equation (\ref{m_eq_1}) will take the form\begin{equation}
m=[c_{1}N_{1}^{-a+1}m+c_{2}N_{1}^{-2a+2}+c_{3}N_{1}^{-2a+1}m+O(m^{2})N_{1}^{-a+1}](1+c_{4}N_{1}^{-a+1}+o(N_{1}^{-a+1}))\label{m_eq1_short}\end{equation}
where $c_{i}=c_{i}(N)$ tend, as $N\to\infty$, to nonzero constants
$d_{i}$, from which we will need only \[
c_{1}=-\frac{F_{1}}{2}\frac{N_{1}}{N_{1}+1}(\frac{\hat{M}_{1}}{N_{1}+1})^{a-1}\to d_{1}=-\frac{F_{1}}{2}M_{1}^{a-1}\]
 \[
c_{4}=-\frac{a^{-1}F_{1}}{2}N_{1}(\frac{\hat{M}_{1}}{N_{1}+1})^{a}\to d_{4}=-\frac{a^{-1}F_{1}}{2}M_{1}^{a}\]
as $\hat{M}_{1}\to M_{1}$ (note that $c_{1},c_{4}$ are obtained
only from the first terms of the expansion). It is evident that the
equation (\ref{m_eq1_short}) has the unique solution $m=o(1)$, which
is asymptotically equal to \begin{equation}
m\sim d_{2}N_{1}^{-2a+2}\label{asymp_m_1}\end{equation}
At the same time, as it can be seen from the comparison of the expansions
(\ref{Exp_delta1}) and (\ref{Exp_delta_N-1}), $\delta_{1,N_{1}+1}-\delta_{2,N_{2}+1}$
look similarly, but have minus sign in front of $c_{1}$ and of $c_{4}$.
Subtracting the second equation from the first we get\[
0=2c_{1}N_{1}^{-a+1}m+2c_{4}c_{2}N_{1}^{-3a+3}+2c_{4}c_{3}N_{1}^{-3a+2}m+O(m^{2})N_{1}^{-a+1}\]
 from where \begin{equation}
m\sim-\frac{2c_{4}c_{2}}{c_{1}}N_{1}^{-2a+2}\sim-\frac{2d_{4}d_{2}}{d_{1}}N_{1}^{-2a+2}\label{asymp_m_2}\end{equation}
 Comparing (\ref{asymp_m_1}) and (\ref{asymp_m_2}) we get the necessary
compatibility condition of the two equations \[
-\frac{2d_{4}}{d_{1}}=2a^{-1}M_{1}=1\]
 But as $M_{1}$ and $M_{2}$ are completely symmetric, we could perform
the same calculations for $M_{2}$. At the end of these calculations
we get the similar condition \[
2a^{-1}M_{2}=1\]
However, these conditions cannot hold simultaneously as $\gamma\neq1$
and thus $M_{1}\neq M_{2}$. Remind that the case $M_{1}=M_{2}$ was
considered separately above. This proves the first assertion of theorem
3.

The proof of the last assertion fo theorem 3 is sufficiently simple.
Let us construct two fixed configurations on the intervals $[0,M]$
and $[M,L]$, as the section 2.3. To get a fixed configuration on
the circle, the particles at the points $0$ и $M$ should be in equilibrium.
For this to happen it is sufficient to adjust the value of the external
force at these points so that it compensated the difference of the
forces from the neighbor particles.

\section{Remarks}

Earlier the ground states (fixed configurations) of classical particles
were studied in connection with the problem of existence and the structure
of the lattice for the condensed matter state (see \cite{Ventevogel_1},
\cite{Duneau_01},\cite{Radin_01}-\cite{Radin_04}). There are also
other continuum one-dimensional models, most known are the Toda chains
and the Frenkel-Kontorova model. In all these models it is assumed
that the potential $V$ has a minimum. Thus in the Frenkel-Kontorova
model mainly the quadratic hamiltonian is considered \cite{BraunKiv},
but there are papers where the latter model is understood in a wider
sense, see for example \cite{Zaslavskij,Gambaudo}. However the ground
states are considered on the whole real line. In our paper we pursue
completely different goals and study different phenomena in a finite
volume, related to the appearance of a finer scale, that is in fact
related to the second term of the asymptotics of distances between
particles. We use direct approach, which is rather straightforward
ideologically, but demands cumbersome calculations.

\end{document}